\documentclass[default]{aastex631}

\usepackage{graphicx}
\usepackage{natbib}
\usepackage{multirow}
\usepackage{makecell}
\usepackage{mwe}

\usepackage{hyperref}

\received{October 6, 2023}
\accepted{August 16, 2024}

\begin{document}

\title{Modeling \emph{Spitzer} 3.6 and 4.5 $\mu$m Eclipse Depths for the Inflated Hot Jupiter in the Evolved Binary System HD 202772}

\correspondingauthor{Arthur D. Adams}
\email{arthur@virginia.edu}
\author[0000-0002-7139-3695]{Arthur D. Adams}
\affiliation{Department of Astronomy, University of Virginia, 530 McCormick Rd, Charlottesville, VA 22904}
\affiliation{Department of Earth and Planetary Sciences, University of California, Riverside, CA 92521, USA}

\author[0000-0002-4420-0560]{Kimberly Bott}
\affiliation{Department of Earth and Planetary Sciences, University of California, Riverside, CA 92521, USA}

\author[0000-0002-4297-5506]{Paul A. Dalba}
\altaffiliation{Heising-Simons 51 Pegasi b Postdoctoral Fellow}
\affiliation{Department of Astronomy and Astrophysics, University of California, Santa Cruz, CA 95064, USA}
\affiliation{SETI Institute, Carl Sagan Center, 339 Bernardo Ave, Suite 200, Mountain View, CA 94043, USA}

\author[0000-0002-3551-279X]{Tara Fetherolf}
\affiliation{Department of Physics, California State University, San Marcos, CA 92096, USA}
\affiliation{Department of Earth and Planetary Sciences, University of California, Riverside, CA 92521, USA}

\author[0000-0002-7084-0529]{Stephen R. Kane}
\affiliation{Department of Earth and Planetary Sciences, University of California, Riverside, CA 92521, USA}

\author[0000-0002-1835-1891]{Ian Crossfield}
\affiliation{Department of Physics and Astronomy, University of Kansas, 1251 Wescoe Dr, Lawrence, KS 66045, USA}

\author[0000-0001-5727-4094]{Drake Deming}
\affiliation{Department of Astronomy, University of Maryland at College Park, College Park, MD 20740, USA}

\author[0000-0003-2313-467X]{Diana Dragomir}
\affiliation{Department of Physics and Astronomy, University of New Mexico, 210 Yale Blvd, Albuquerque, NM 87131, USA}

\author[0000-0002-8990-2101]{Varoujan Gorjian}
\affiliation{Jet Propulsion Laboratory,
California Institute of Technology, 4800 Oak Grove Drive, Pasadena, CA 91109, USA}

\author[0000-0003-0514-1147]{Laura Kreidberg}
\affiliation{Max-Planck-Institut für Astronomie, Königstuhl 17, D-69117 Heidelberg, Germany}

\author[0000-0001-9414-3851]{Farisa Y. Morales}
\affiliation{Jet Propulsion Laboratory,
California Institute of Technology, 4800 Oak Grove Drive, Pasadena, CA 91109, USA}

\author[0000-0003-4990-189X]{Michael W. Werner}
\affiliation{Jet Propulsion Laboratory,
California Institute of Technology, 4800 Oak Grove Drive, Pasadena, CA 91109, USA}

\begin{abstract}

As an inflated Hot Jupiter orbiting an early-type primary star in the evolved binary HD 202772 system, HD 202772 A b's presence invites a study of how such a planet forms and evolves. As a prelude to potential atmospheric characterization with the latest generation of observatories, we present a reduction and analysis of eclipse light curve observations of HD 202772 A b acquired with the \emph{Spitzer} Space Telescope using the 3.6 and 4.5 $\mu$m channels. We find eclipse depths of $680\pm68$ and $1081^{+54}_{-53}$ ppm, respectively, corresponding to day-side effective temperatures of $2130^{+102}_{-91}$ and $2611^{+46}_{-49}$ K. The corresponding Bond albedos are consistent with the distribution of albedos for Hot Jupiters observed with both \emph{Spitzer} and \emph{TESS} The heat redistribution efficiencies consistent with the Bond albedo range predicted by 1-D atmospheric models in radiative-convective equilibrium are $0.71\pm0.10$ and $0.03^{+0.03}_{-0.02}$, respectively, indicating a weak day-night contrast for the former and a strong contrast for the latter. Given this, and the unique environment in which this planet resides, we recommend follow-up observations with \emph{JWST} to more precisely constrain its atmospheric composition and structure, as well as its host stellar environment, to elucidate if and how the atmospheres of these close-in giants evolve with host stars in binaries past the main sequence.

\end{abstract}

\section{Introduction} \label{sec:introduction}
The Transiting Exoplanet Survey Satellite (\emph{TESS}) has observed many planet-hosting stars, including those that have evolved off the main sequence. Their planetary systems provide a growing sample of close-in gas giants, refining the picture of how these planets evolve at the late-to-end stages of their hosts. Prior to the launch of \emph{TESS}, there was considerable uncertainty as to whether planets occurred less frequently around evolved hosts, as well as those in binary systems \citep{Wang2015}; both environmental conditions are predicted to have a significant effect on the formation and evolution of planetary systems \citep[see e.g.][]{fab07,Veras2016}. The occurrence rate of Hot Jupiters should be dependent on stellar age, as planets are susceptible to orbital in-spiral \citep[e.g.][]{Vissapragada2022} and eventual engulfment, with an estimated 70\% of all gas giants meeting this fate if they orbit an A-type host \citep{Stephan2018}. Considering stellar multiplicity, \citet{Fontanive2021} find that close-in giants appear to be more frequent around the most massive stars in multiple star systems than they are around the general population of planet hosts. 

HD 202772 (also known as TOI-123) is an example of both environments, consisting of a binary star with the primary, having evolved off the main sequence to become a subgiant, hosting a giant planet. HD 202772 A b is an inflated Hot Jupiter with a mass of $1.017^{+0.070}_{-0.068}$ $M_\mathrm{J}$ and a radius of $1.545^{+0.052}_{-0.060}$ $R_\mathrm{J}$, orbiting its host star with a period of approximately 3.3 days \citep{Wang2019a}. As HD 202772 A is an F-type subgiant with an effective temperature of $6272^{+77}_{-71}$ K, the planet receives an instellation flux nearly 3500 times that of Earth, yielding an estimated equilibrium temperature of $2132^{+28}_{-30}$ K (as calculated via the equations in \citealp{knu09b,Kane2011a}). Planets experiencing this much irradiation from their hosts are a major focus of study, as the mechanisms that drive radius inflation also fundamentally drive these planets' atmospheric structures and circulation. At a distance of 161 parsecs, the host is also bright ($m_J=7.232\pm0.026$) which, when combined with the short orbital period and large scale height of the planet's atmosphere, makes this system a strong candidate for atmospheric characterization with \emph{JWST}. To this point, \citet{Wang2019a} point to the Transmission Spectroscopy Metric \citep{Kempton2018,Louie2018}, which provides a proxy for a S/N estimate for a planet in transmission based on fundamental system parameters. The estimated TSM value for HD 202772 A b is approximately 210 for a scale factor of 1.15, though the original TSM is only defined for planets up to 10 $R_\oplus$ and for hosts as bright as $m_\mathrm{J} \approx 9$.

Atmospheric characterization efforts rely on a combination of directly resolved imaging, eclipse and transit observations, and time-resolved phase curves. Where eclipsing planets were previously limited to photometry or low-resolution spectroscopy, we now have the capability for high resolutions ($R\gtrsim 10^4$--$10^5$) across all observing types, both from the ground and in space. A recent example is the $R\sim10^5$ pre-eclipse phase curve spectrum of the ultra-hot Jupiter MASCARA-1 b \citep{Ramkumar2023}. As a point of comparison to HD 202772 A b, the similarly inflated Hot Jupiter WASP-76 b, orbiting a similarly bright F-type host, has been a prime target for recent observations and modeling efforts \citep[e.g.][]{Garhart2019,May2021,Beltz2022}. HD 202772 A b's estimated equilibrium temperature places it in a valuable position within the emerging spectral sequence of highly-irradiated gas giants. \citet{Mansfield2020a} have found that, as day-side temperatures reach approximately 2100 K, Hot Jupiters are expected to rapidly form thermal inversions in their upper atmospheres, which switches the resulting spectrum from being dominated by absorption features to those in emission. This is believed to be a consequence of the onset of TiO and VO in the gas phase, as discussed in works such as \citet{Fortney2008}. The expected abundance and composition of clouds from microphysical models also exhibits a major change around 2100~K, with cooler Hot Jupiters expected to have strong contributions from silicate, iron, and corundum condensates while those a bit hotter will not have those species condense into clouds, and are therefore predicted to be relatively cloud-free \citep{Gao2020,Zhang2020}.

With the backdrop of current observations and characterizations of this class of giant planets, one the most foundational datasets to this sub-field of exoplanet science remain thermal eclipse depths. The pioneering telescope for eclipse depths was the \emph{Spitzer} Space Telescope, launched in 2003 and remaining operational through early 2020 \citep[see e.g.][among many others]{wer04,Deming2020,Scire2022}. The data we present here come from a ``post-cryogenic'' phase of Spitzer, when only the 3.6 and 4.5 $\mu$m channels of the Infrared Array Camera (IRAC) were still operational. This work is an expansion of one of the planets included in the re-analysis of all Spitzer secondary eclipse targets in \citet{Deming2023}. We will describe the technique of reducing the data, including identifying and subtracting out now well-modeled noise patterns, assessing the use of an additional noise model (such as Gaussian process regression) in cases where the existing models fail to capture most of the observed correlated noise, then using a well-established simple model to place HD 202772 A b's eclipse measurements in context of a wealth of planets previously observed and characterized in eclipse.

\section{Observations} \label{sec:observations}
\citet{Wang2019a} confirmed the planetary status of HD 202772 A b with photometry from the Transiting Exoplanet Survey Satellite (\emph{TESS}), with observations taken during Sector 1 of observations. The authors complemented the \emph{TESS} data with novel observations of the system using a multitude of ground-based instruments, including images at $0\farcs05$ spatial resolution with adaptive optics using the NIRC2 instrument at the Keck Observatory, as well as optical spectroscopy with the HIRES instrument; and optical spectra with the Network of Echelle Spectrographs (NRES) at the Las Cumbres Observatory (LCO), as well as with CHIRON at the Cerro Tololo telescope and the Tillinghast Reflector Echelle Spectrograph (TRES) at the Fred L. Whipple Observatory (FLWO). \citet{Wang2019a} use these observations in tandem to further confirm the binarity of the system (with HD 202772 B as a companion at a separation of approximately $1\farcs3$), place joint constraints on the stellar parameters, and provide radial velocity measurements of HD 202772 A b. HD 202772 is therefore a well-characterized system, with planet Ab having constraints of $\sim 5$\% in mass and radius, and eclipse ephemerides within $\sim \pm 1$ hour, and our work benefits from their global analysis.

HD 202772 A b was observed with \emph{Spitzer} in secondary eclipse in the 3.6 and 4.5 $\mu$m IRAC channels, each used on separate observations spanning 10.35 and 12.13 hours, on 2019 March 2--3 (BJD 2458545) and 2019 February 14 (BJD 2458528--2458529), respectively. These observations were part of the \emph{Spitzer} GI Program \#14084 (PI: Ian Crossfield). From these observations we obtain the basic calibrated data product, which consists of 64 frames of $32\times32$ pixels each, with a frame time of 0.4 seconds.

\begin{deluxetable*}{lc}\label{table:HD202772A_parameters}
\tabletypesize{\scriptsize}
\tablewidth{0pt}
\tablecaption{Selected parameters of the HD 202772 A system from \citet{Wang2019a}.}
\tablehead{\colhead{Parameter} &
           \colhead{Value}
           }
\startdata
$R_\star/R_\odot$           & $2.591^{+0.078}_{-0.093}$ \\
$T_\star$ (K)               & $6272^{+77}_{-71}$        \\
$R_\mathrm{p}/R_\mathrm{J}$ & $1.545^{+0.052}_{-0.060}$ \\
$T_\mathrm{eq}$ (K)          & $2132^{+28}_{-30}$        \\
$a/R_\star$                 & $4.33^{+0.15}_{-0.13}$    \\     
\enddata
\end{deluxetable*}

\section{Modeling} \label{sec:model}
\subsection{Eclipse Parameters}\label{sec:model:eclipse}
We use these \emph{Spitzer} observations to constrain the disk-integrated day-side emission of HD 202772 A b, which can then informs models about the planet's atmosphere. Day-side emission fluxes correspond to eclipse depth, and in order to obtain estimates of the eclipse depths in each \emph{Spitzer} channel, we must first reduce the data and isolate the eclipse signal from systematic effects from the instrument. The reduction begins with the observation frames; for each frame, we
\begin{enumerate}
    \item remove 4-$\sigma$ outliers in the temporal dimension for each pixel,
    \item perform background subtraction by masking the central $4\times4$ pixels and subtracting the median count value of the remaining frame,
    \item calculating the centroid of the target star using \texttt{photutils} \citep{Bradley2023}, a software package in Python, and
    \item calculate the flux in 12 apertures, each centered on the centroid and spanning 1.25--4 pixels in increments of 0.25 pixel.
\end{enumerate}
To isolate the eclipse signal we use pixel-level decorrelation as originally described in \citet{Deming2015}, and adapted as described in \S 2.1 of \citet{Jontof-Hutter2022}. The goal of this approach is to capture the astrophysical signal more effectively by treating each pixel registering a response from the source as an independent contribution of signal and noise. The aperture of the telescope spreads the light from a point source such as a star onto multiple pixels of the detector, whose contributions to the measured flux time series are modeled as individual ``basis functions''. The total measurement is composed of the model for the astrophysical signal, i.e.~the eclipse curve; a ``ramp'' that captures the time-dependent variation in the response of the instrument, often modeled as a linear or quadratic function of time; and the aforementioned basis functions, which are the remaining fractional contributions to the residual flux, per integration, of each pixel in the aperture after both the astrophysical and ramp models are subtracted. The total measured signal $S\!\left(t\right)$ is, to linear order (i.e.~ignoring cross-terms in a Taylor expansion),
\begin{equation}\label{eq:PLD}
    S\!\left(t-t_0\right) = E\!\left(t-t_0\right)\frac{F_\mathrm{p}}{F_\star} + R\!\left(t-t_0\right) + \sum\limits^{N_\mathrm{pix}}_{i=1}w_i \hat{P}_i\!\left(t-t_0\right).
\end{equation}
The midpoint of eclipse time is treated as a free parameter $t_0$. $F_\mathrm{p}/F_\star$ is the eclipse depth and $E$ represents the chosen depth-normalized shape function. For our eclipse model we use \texttt{batman}, a software package that models exoplanet transit and eclipse light curves \citep{Kreidberg2015}. We choose to fix all system parameters of the model to the best estimates from \citet{Wang2019a} except for $t_0$ and $F_\mathrm{p}/F_\star$; \texttt{batman} then determines the eclipse shape. $R\!\left(t-t_0\right) \equiv R_0 + R_1\!\left(t-t_0\right)$ is the ramp correction function; we take the function to be zero at time $t_0$, hence $R_0$ is assumed to be zero. We settle on a linear function based on inspection of the residuals. Each pixel (of a total $N_\mathrm{pix}=13$) within the aperture at time $t$ can contribute to the eclipse and ramp signals; when these are subtracted, what is left is uncharacterized ``noise'' which is distributed among each pixel. There is a slight difference between our pixel averaging and that presented in \citet{Deming2023}: we use a 13-pixel pattern,  a diamond 5 pixels wide and tall; \citet{Deming2023} use a 12-pixel pattern representing a $4\times4$ grid with the corner pixels omitted. Since for \emph{Spitzer} the centroid of the star moves on average less than a pixel's width across an entire observation, the same collection of pixels comprise the aperture, but the proportion of flux each pixel receives will change over the duration of the observation. $\hat{P}_i$ is then the fraction of the residual flux captured by pixel $i$. Each pixel's contribution gets a weight $w_i$ in the final model of the flux which models the pixel's inherent efficacy to respond to the astrophysical source. If the noise in each pixel were completely independent, Gaussian (white) noise, we would expect the posterior distributions of each weight to be approximately Gaussian, centered at zero, and with a spread that represents the scale of the noise. Correlated or structured noise will cause the locations and spreads of the distributions to shift, including cross-correlations between pixel weights. The free parameters therefore are $t_0$, $F_\mathrm{p}/F_\star$, the weights $w_i$, and the slope of the ramp function $R_1$.

We use \href{https://emcee.readthedocs.io}{\texttt{emcee}}, a Markov-Chain Monte Carlo parameter estimation tool, to fit the model to the data and estimate the correlations and uncertainties in each parameter \citep{Foreman-Mackey2013}. The chains for the models in each channel used 200 walkers each running for 5000 steps. Each walker begins at $w_i=0$, $R_1=0$, and the initial eclipse depth is estimated using the system parameters from \citet{Wang2019a}. The priors are broad, restricted only by physical limits: namely, that the mid-point of the eclipse occur between the timestamps of the first and last data points, and that the planet-star flux ratio be between 0 and 1. The MCMC run proceeded with 200 chains for 5000 steps, for a total of $10^6$ samplers, and the ``burn-in'' phase was taken to be the first 1500 steps, as judged by inspection of the chains' progress and convergence.

In order to produce the eclipse model for the above analysis, one must choose both an aperture size in pixels and a number of data points per bin (a ``binning factor'') for binning both the target star light curve and basis vector light curves. The goal is to choose a pair of aperture and bin sizes that minimizes the correlated (``red'') noise, which is evaluated as follows. We perform a coarse grid search, with aperture sizes ranging from 1.25--4 pixels by 0.25 pixel and binning factors of 50, 100, 150, and 200. For each pair of aperture and bin size, bin the data and basis light curves accordingly, then
\begin{itemize}
    \item run the above MCMC routine to get a set of best-fit pixel weights and eclipse light curve parameters;
    \item calculate the residuals for each unbinned data point to the model light curve generated from those parameters;
    \item bin those residuals by a series of different bin sizes from 1 (unbinned) to some upper bin size, calculating the relative root-mean-square (RMS) deviation from the model binned equivalently;
    \item compare this function with the expectation for uncorrelated noise, i.e.~where the log of the RMS residuals as a function of the log of the binning factor follows a $-1/2$ slope\footnote{This is also known as the ``Allan deviation relation'' \citep{Allan1966}.};
    \item select the combination of aperture radius and bin size with a residual log-log slope closest to $-1/2$; or,
    \item if the residuals imply that significant correlated residuals exist, apply a Gaussian process regressor to try to account for noise that is correlated in time --- then select the most appropriate closest aperture and bin combination by the above criterion.
\end{itemize}

We employ a simple Gaussian process regression model to capture any residual noise in the eclipse data that is correlated in time after accounting for pixel-to-pixel correlation. (For key references in implementing Gaussian processes, see e.g.~\citealt{OHagan2006} and \citealt{books/lib/RasmussenW06}.) To build our emulators, we use the \texttt{scikit-learn} package in Python \citep{scikit-learn}. We use a radial basis function which only depends on the time interval between two points,
\begin{equation}\label{eq:RBF_kernel}
k\!\left(t_a, t_b\right) = \exp\left[-\frac{\left(t_{a}-t_{b}\right)^2}{2\tau^2}\right],
\end{equation}
where $\tau$ represents the characteristic time interval over which the noise model is correlated; and a white kernel function
\begin{equation}\label{eq:white_kernel}
k\!\left(t_a, t_b\right) = \sigma \delta_{ab}.
\end{equation}
We construct our regressor with this combined kernel and use it to fit the noise profile of each channel's unbinned residual time series.

For Channel 1 the best fit aperture size and binning factor are 2.25 pixels and 150, respectively; for Channel 2  these are 3.75 pixels and a factor of 100. Our residuals are discussed more in \S \ref{sec:results:eclipse-depths}. This approach is comparable to the grid-like search procedure as described in \citet{Deming2015}, but is coarser than the fitting procedure described in \citet{Deming2023}, which uses a multi-variate linear regression. This difference, in addition to the slight difference in pixel aperture shape, may contribute to the differences in our retrieved eclipse depths; this is discussed further in \S \ref{sec:results:eclipse-depths}.

The final step is to incorporate the effects of flux contamination within the aperture from neighboring stars. Since HD 202772 is a binary with a projected angular separation of $1\farcs3$, this contamination is expected to be quite large. We adopt the procedure of \citet{Deming2023}, in their Appendix B: that companion stars closer in separation than $2''$ are taken to have nearly all of their flux within the aperture, and therefore the fluxes may simply be added. We accordingly use their derived dilution factor of 1.207 for both \emph{Spitzer} bands.

\subsection{Astrophysical Parameters}\label{sec:model:astrophysical}
Once the eclipse depths are calculated, we compare the eclipse depths to the corresponding brightness temperature that would yield such an eclipse depth. The total planet-to-star flux ratio depends on
\begin{itemize}
    \item the emission flux, calculated via
    \begin{equation}\label{eq:emission_flux}
        \left(\frac{F_\mathrm{p}}{F_\star}\right)_\mathrm{emit} = \left(\frac{R_\mathrm{p}}{R_\star}\right)^2 \int B_\lambda\!\left(T_\mathrm{p}, \lambda\right)f_\mathrm{Spitzer}\!\left(\lambda\right) \, d\lambda \Bigg/ \int I_\lambda\!\left(T_\star, \left[\mathrm{Fe}/\mathrm{H}\right], \log g, \lambda\right)f_\mathrm{Spitzer}\!\left(\lambda\right) \, d\lambda
    \end{equation}
    where $B_\lambda\!\left(T, \lambda\right) = 2hc^2/\left\{\lambda^5 \left[\exp\left(hc/\lambda k_\mathrm{B} T\right)-1\right]\right\}$ is the Planck function, $I$ is the spectral intensity of the star, and $f_\mathrm{Spitzer}$ is the bandpass response of each channel; and
    \item the reflected light contribution, which is a function of the Bond albedo:
    \begin{equation}\label{eq:reflected_flux}
        \left(\frac{F_\mathrm{p}}{F_\star}\right)_\mathrm{refl} = A_\mathrm{B} \left(\frac{R_\mathrm{p}}{a}\right)^2.
    \end{equation}
\end{itemize}

Our stellar source spectrum comes from the Castelli--Kurucz atlas of stellar atmospheres \citep{Castelli2004} as implemented in \texttt{pysynphot} \citep{pysynphot,synphot}. For the temperature used in Equation \ref{eq:emission_flux}, we model the planet's day-side effective temperature $T_\mathrm{d}$ in terms of the Bond albedo $A_\mathrm{B}$ and heat redistribution efficiency $\varepsilon$, as is done in \citet{Cowan2011}, Equation 4. In terms of the planet's expected equilibrium temperature at its sub-stellar point
\begin{equation}\label{eq:equilibrium_temperature}
    T_\mathrm{eq} = T_\mathrm{eff} \sqrt{\frac{R_\star}{2 a}},
\end{equation}
the day-side temperature can be modeled as
\begin{equation}\label{eq:dayside_temperature}
    T_\mathrm{d} = T_\mathrm{eq} \left[4\left(1-A_\mathrm{B}\right)\left(\frac{2}{3}-\frac{5}{12}\varepsilon\right)\right]^{1/4}.
\end{equation}
$\varepsilon$ is a parameter which ranges from 0 (where all insolation energy is retained on the day side) to 1 (perfectly uniform heat redistribution). Since we are fitting our eclipse depths to the corresponding day-side temperature, constraints on $A_\mathrm{B}$ and $\varepsilon$ are completely degenerate. In \citet{Cowan2011}, a combination of both day- and night-side temperatures are needed to constrain each parameter separately. However, it is still useful to understand the joint constraints on these parameters, and we can use independent models to estimate the effective Bond albedo if the atmosphere were in radiative-convective equilibrium (see \S \ref{sec:model:rad-con}).

The brightness temperature $T_\mathrm{b}$ is a separate measurement of the flux in each band, and is defined as the temperature of a blackbody whose flux matches the observed flux. We borrow the formula for estimating $T_\mathrm{b}$ from Equation 1 of \citet{Baxter2020},
\begin{equation}\label{eq:brightness_temperature}
    T_\mathrm{b} = \frac{h c}{k_\mathrm{b} \lambda} \left[\ln\!\left(\frac{2 h c^2 \pi \delta_\mathrm{tra}}{\lambda^5 \bar{F}_\mathrm{\star}\!\left(\lambda\right) \delta_\mathrm{occ}}\right)\right]^{-1}
\end{equation}
where $c$ is the speed of light; $h$ and $k_\mathrm{b}$ are the Planck and Boltzmann constants, respectively; $\lambda$ is the central wavelength of a spectral element in the channel; $\delta_\mathrm{tra} = \left(R_\mathrm{p}/R_\star\right)^2$ is the transit depth and $\delta_\mathrm{occ} = F_\mathrm{p}/F_\star$ our measured eclipse depth in the given channel; and $\bar{F}_\mathrm{\star}\!\left(\lambda\right)$ is the stellar flux integrated over the same channel's response function. The observed brightness temperature will be an average of brightness temperatures calculated as above at individual wavelengths, weighted by the response function of each \emph{Spitzer} bandpass.

To estimate the uncertainties on the fitted day-side temperature we sample the $A_\mathrm{B}$--$\varepsilon$ parameter space using nested sampling, via the \texttt{dynesty} package in Python \citep{Speagle2020}. We use 16 live points, sample with the uniform sampler with multiple (``multi''), and run until the default stopping criterion, which is designed to adequately assess both the evidence and posterior distributions. The primary goal is to get reliable uncertainties in these parameters (and correspondingly, the day-side temperature) by estimating the posterior distributions of $A_\mathrm{B}$ and $\varepsilon$, and accordingly our priors are as wide as is physically plausible (both range from 0 to 1).

\subsection{Radiative-Convective Models}\label{sec:model:rad-con}
We use an atmosphere and climate model \citep[\texttt{PICASO 3.0}, see][and the \href{https://natashabatalha.github.io/picaso/notebooks/climate/12b_Exoplanet.html}{PICASO documentation website}]{Batalha2019,Mukherjee2023} to compare the eclipse depths with those predicted in radiative-convective equilibrium. 1-D atmospheric models have also been employed in works such as \citet{Baxter2020} and \citet{Deming2023} to place the inferred properties of the \emph{Spitzer} Hot Jupiter sample in context of trends in metallicity and C/O ratio. Since we focus on a single system, we are able to fix the stellar parameters and comment more directly on the specific atmospheric chemistry that most closely matches HD 202772 A b. We run 9 models, corresponding to each pair of C/O ratios and metallicities chosen from 0.5, 1, and 2 $\times$ the solar C/O ratio, and [Fe/H] $=-1$, 0, and 1, respectively. The opacity data for these chemistry configurations in \texttt{PICASO} is drawn from \citet{Lupu2023}. The climate code starts with an initial set of planet and instellation parameters, including the temperature-pressure profile, and iterates until a solution approximating radiative-convective equilibrium. We fix our models to one convective zone and compute the spectrum for the limiting case of a day side on a tidally locked planet, and use the system properties given in Table \ref{table:HD202772A_parameters}, along with the derived surface gravity $\log g = 3.02$. We then generate emission spectra in the relevant wavelength ranges, and use our existing pipeline to convert these model spectra into eclipse depths in the warm \emph{Spitzer} bands. Note that the uncertainties in our modeled eclipse depths are based on the observed system parameters alone and should be considered lower limits. Each radiative-convective model solution produces an effective temperature; by plugging this in as the day-side temperature in Equation \ref{eq:dayside_temperature}, we can calculate the effective Bond albedo and then estimate the heat redistribution efficiency under the physical assumptions of our models.

\section{Results} \label{sec:results}
\subsection{Eclipse Depths}\label{sec:results:eclipse-depths}
\begin{deluxetable*}{rcccc}\label{table:PLD_parameters}
\tabletypesize{\footnotesize}
\tablewidth{0pt}
\tablecaption{Median and maximum likelihood estimator (MLE) parameter values for the parameters of the pixel-level decorrelation model fit to the eclipse light curves of HD 202772 A b, for \emph{Spitzer} IRAC Channels 1 (centered at 3.6 $\mu$m) and 2 (centered at 4.5 $\mu$m). $w_i$ is the weight of the $i$-th pixel used in the aperture for the pixel-level decorrelation, and $R_1$ is the slope of the linear term in the eclipse model; both are described in \S \ref{sec:model:eclipse}. Note that the eclipse depths reported here are not yet corrected for flux dilution from the secondary star.}
\tablehead{
    {} &
    \multicolumn{2}{c}{3.6 $\mu$m} &
    \multicolumn{2}{c}{4.5 $\mu$m} \\
    \colhead{Parameter} &
    \colhead{68\% Confidence Interval} &
    \colhead{MLE} &
    \colhead{68\% Confidence Interval} &
    \colhead{MLE}
    }
\startdata
$t_0$ (BJD$-2458528$) & $17.4100^{+0.0018}_{-0.0017}$ & $17.4172$ & $0.8700^{+0.0011}_{-0.0012}$ & $0.8664$ \\
$F_\mathrm{p}/F_\star$ (ppm) & $563^{+56}_{-57}$ & $452$ & $895\pm44$ & $1004$ \\
$w_{0}$ & $0.97\pm0.45$ & $1.50$ & $-0.08\pm0.20$ & $0.59$ \\
$w_{1}$ & $-1.42\pm0.32$ & $-2.44$ & $-0.63\pm0.14$ & $-0.79$ \\
$w_{2}$ & $0.19\pm0.14$ & $0.47$ & $-0.11\pm0.08$ & $-0.18$ \\
$w_{3}$ & $-1.01\pm0.15$ & $-0.96$ & $0.31^{+0.10}_{-0.11}$ & $0.20$ \\
$w_{4}$ & $1.05\pm0.48$ & $0.55$ & $0.05^{+0.21}_{-0.22}$ & $-0.47$ \\
$w_{5}$ & $-0.17\pm0.05$ & $-0.06$ & $0.11\pm0.02$ & $0.13$ \\
$w_{6}$ & $0.34^{+0.05}_{-0.04}$ & $0.44$ & $0.01\pm0.02$ & $0.03$ \\
$w_{7}$ & $0.43^{+0.24}_{-0.23}$ & $-0.08$ & $-0.13\pm0.04$ & $-0.08$ \\
$w_{8}$ & $-0.45^{+0.46}_{-0.45}$ & $-1.72$ & $0.43\pm0.20$ & $0.22$ \\
$w_{9}$ & $-0.18\pm0.10$ & $-0.18$ & $-0.09\pm0.04$ & $-0.11$ \\
$w_{10}$ & $-0.66\pm0.12$ & $-0.50$ & $0.06^{+0.02}_{-0.01}$ & $0.06$ \\
$w_{11}$ & $-0.49^{+0.20}_{-0.21}$ & $-0.44$ & $-0.02^{+0.09}_{-0.08}$ & $0.02$ \\
$w_{12}$ & $2.36\pm0.61$ & $2.23$ & $-0.22\pm0.26$ & $-0.16$ \\
$R_1$ & $0.0048^{+0.00064}_{-0.00062}$ & $0.0051$ & $-0.00014^{+0.00039}_{-0.0004}$ & $-0.00056$ \\
\enddata
\end{deluxetable*}

\begin{figure*}[htb!]
\begin{center}
\includegraphics[width=17cm]{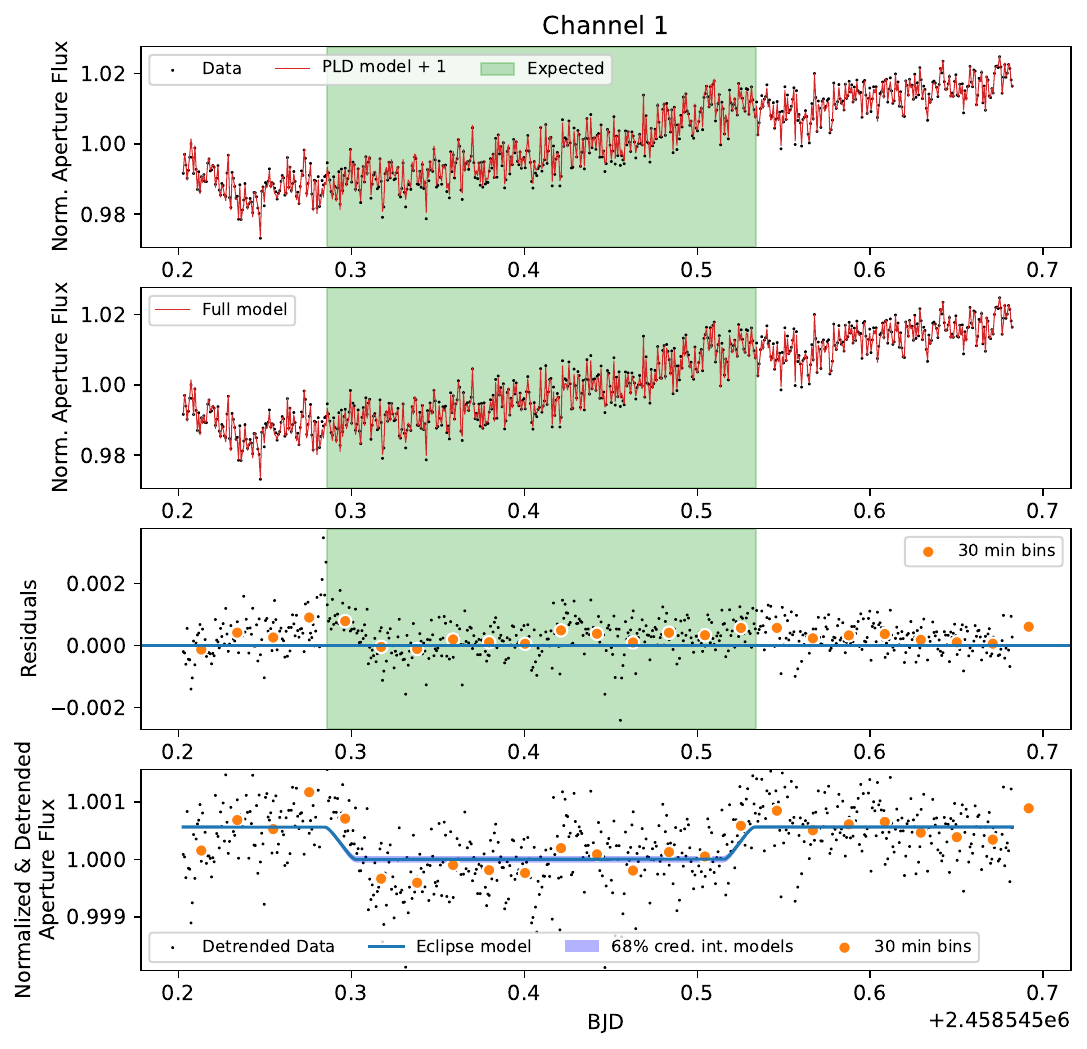}
\caption{The setup and fits to the pixel-level decorrelation (PLD) technique used to reduce the \emph{Spitzer} data for observations of HD 202772 A in the channel centered at 3.6 $\mu$m. The raw light curve is shown in black on the top 2 sub-plots, with the model fit from PLD in red. The green shaded region shows when we predict in time the planet will be in secondary eclipse. The lower 2 sub-plots show the residuals and de-trended eclipse light curve after accounting for systematics via PLD. The data points in orange are binned to 30-minute intervals, with the 68\% confidence interval of the binned and de-trended light curve in blue.}
\label{fig:PLD_ch1}
\end{center}
\end{figure*}

\begin{figure*}[htb!]
\begin{center}
\includegraphics[width=17cm]{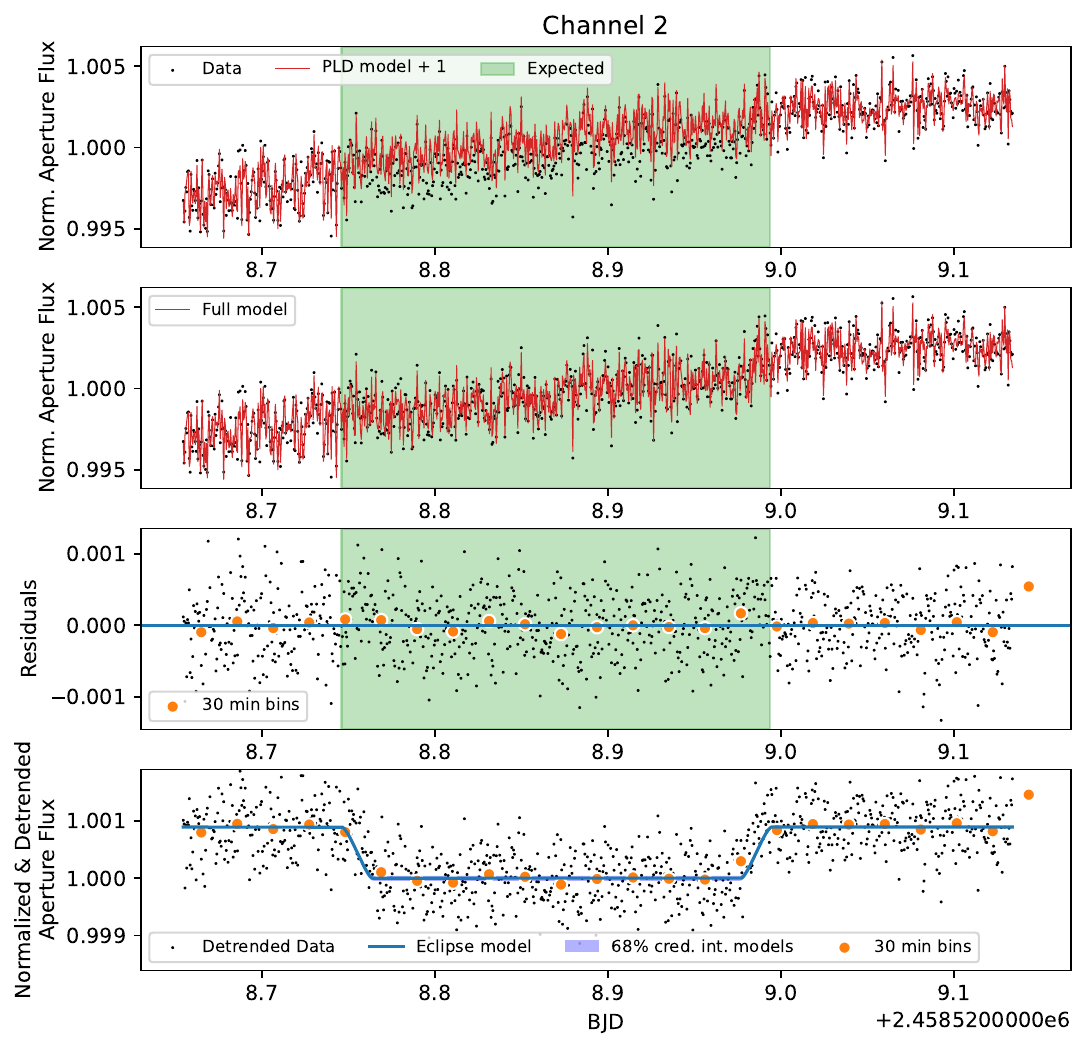}
\caption{The setup and fits to the pixel-level decorrelation (PLD) technique used to reduce the \emph{Spitzer} data for observations of HD 202772 A in the channel centered at 4.5 $\mu$m. The raw light curve is shown in black on the top 2 sub-plots, with the model fit from PLD in red. The green shaded region shows when we predict in time the planet will be in secondary eclipse. The lower 2 sub-plots show the residuals and de-trended eclipse light curve after accounting for systematics via PLD. The data points in orange are binned to 30-minute intervals, with the 68\% confidence interval of the binned and de-trended light curve in blue.}
\label{fig:PLD_ch2}
\end{center}
\end{figure*}

The results of the pixel-level decorrelation (PLD) on the \emph{Spitzer} photometry for HD 202772 A b are shown in Figure \ref{fig:PLD_ch1} (Channel 1, centered at 3.6 $\mu$m) and Figure \ref{fig:PLD_ch2} (Channel 2, centered at 4.5 $\mu$m). All retrieved posterior distributions are approximately Gaussian, with a handful of significantly correlated distributions between the weights of individual pixels. This is expected since the response of pixels within an aperture, especially neighboring pixels, may be correlated as a function of the centroid position. Some of the basis functions corresponding to pixel time series show linear trends in time. Several of the posterior distributions are significantly offset from zero, implying that correlated noise is contributing to the fit. Additionally, the two channels share some cross-correlations between pairs of pixel weights; since the system was observed in each channel on separate observing runs, this implies that a considerable amount of the correlated noise is due to systematic errors in the same set of pixels. Note that the correlated errors here refer to those accounted for in the pixel-level decorrelation step, rather than any residual correlated noise remaining after de-trending.

\begin{figure}[htb!]
    \begin{center}
        \begin{tabular}{cc}
            \includegraphics[width=0.475\textwidth]{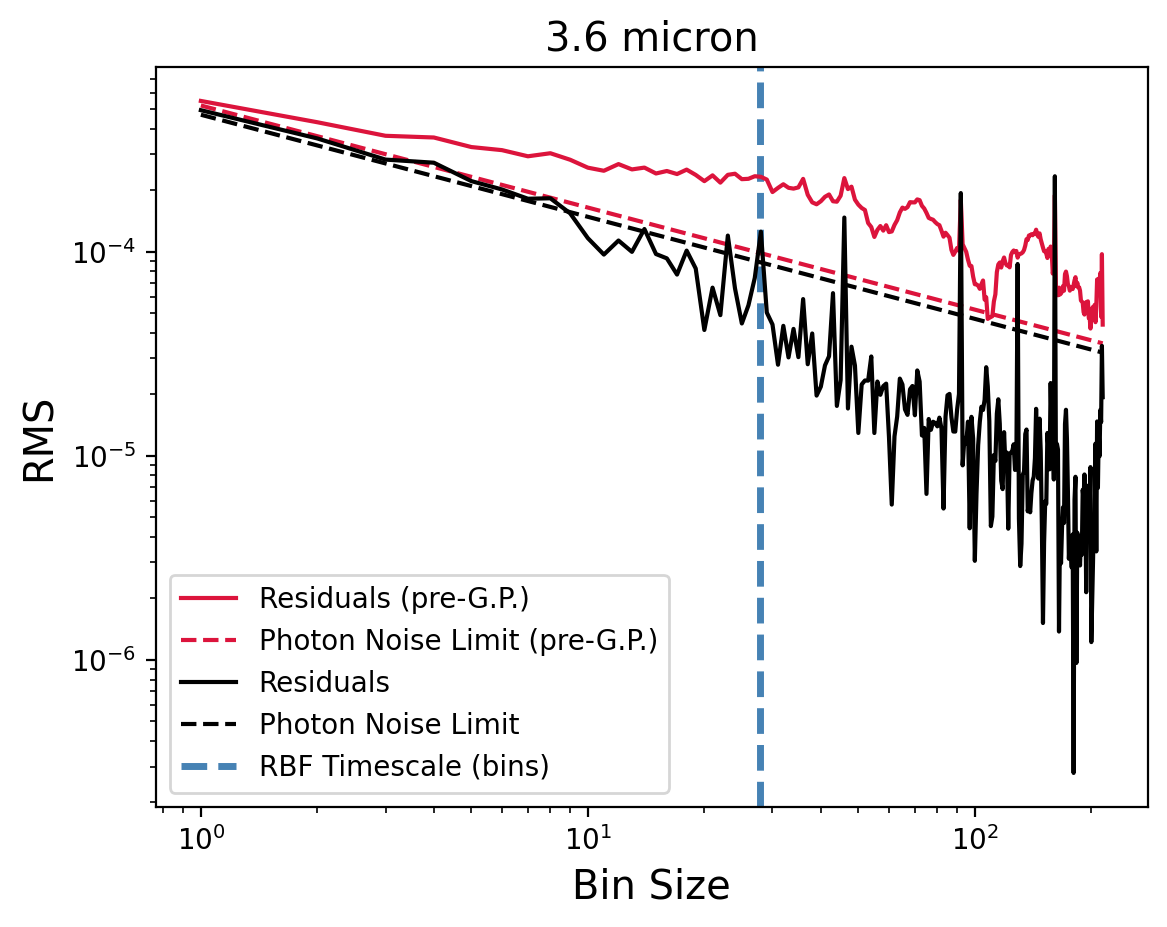} & \includegraphics[width=0.475\textwidth]{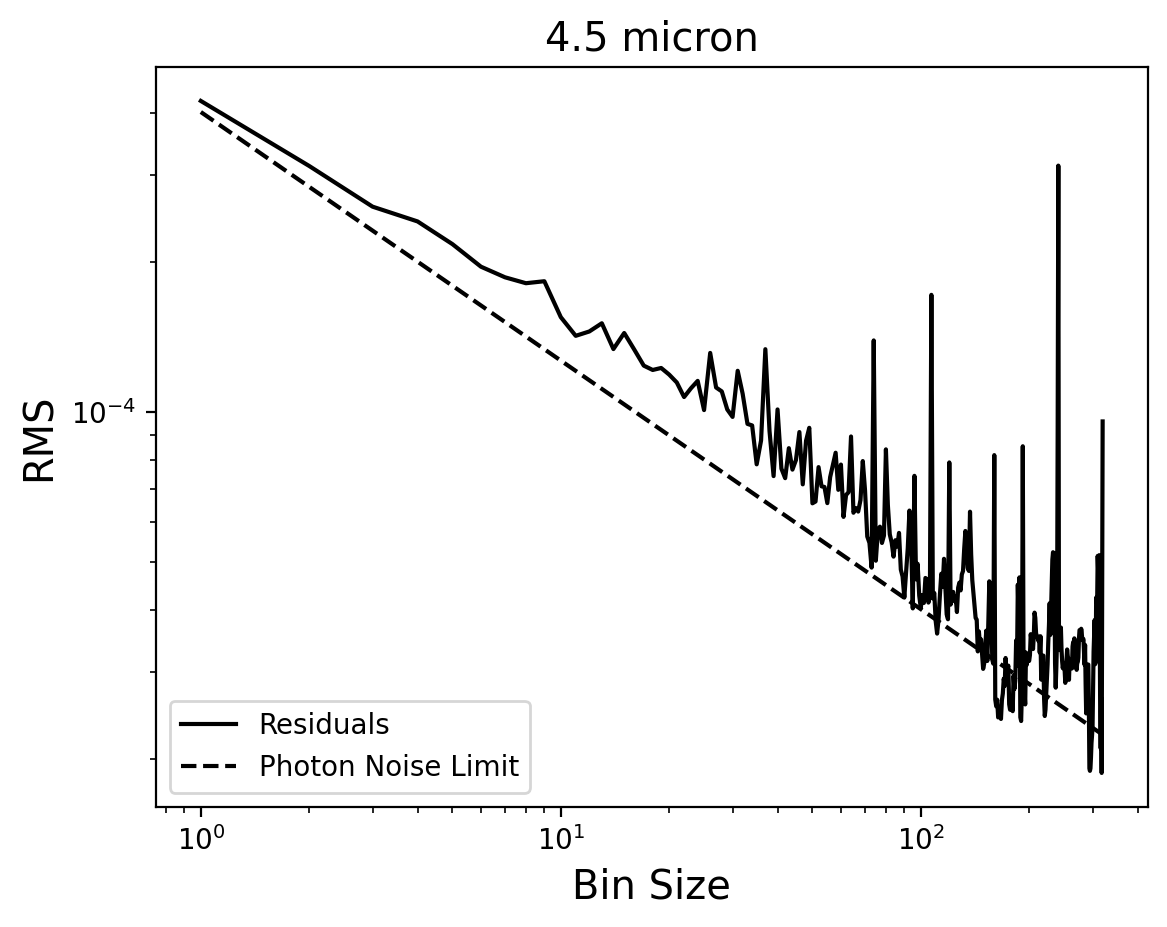} \\
        \end{tabular}
    \end{center}
    \caption{The root-mean-square (RMS) residuals in our eclipse model fits to the data, as a function of the bin size used in the process described in \S \ref{sec:model:eclipse}. The solid line represents the residuals, and the dashed line represents the ``ideal'' expectation of pure photon noise (i.e. no correlated noise), which should follow a $-1/2$ slope in the log-log space of log bin size versus RMS residuals. For the 3.6 $\mu$m channel, a Gaussian process regression is added to the original (red) noise profile, yielding the result shown in black. The time scale of the radial basis function ($\tau \approx 29.7$ minutes $= 27.8$ bins in Equation \ref{eq:RBF_kernel}) is shown as a vertical blue dashed line. One caveat of this approach is that the final fit suffers from some over-subtraction of noise at bin sizes longer than the time scale $\tau$, as the kernel function becomes under-sampled.}
    \label{fig:PLD_residuals}
\end{figure}

Applying the Gaussian process regressor to the PLD-subtracted residuals, Channel 1 (3.6 $\mu$m) returned a time scale $\tau \approx 30$ minutes, which corresponds to a bin size of $\approx 28$ data points. Channel 2 (4.5 $\mu$m) returned a time scale $\approx 8$ times the full duration of observations, which indicates there is not a significant time-correlated noise signal. Therefore, we only apply the Gaussian process regression on the Channel 1 data, with a best-fit white kernel noise scale $\sigma \approx 0.83$ times the original variance in the unbinned residuals. This corresponds to a typical error bar of $\approx 40$ parts per million in the bins that are used for the astrophysical models. Figure \ref{fig:PLD_residuals} show these residuals compared with the expectation value of the ``photon'' noise limit (corresponding to completely uncorrelated noise). Note that, while the Gaussian process regression solves the excess noise, it over-subtracts noise for bin sizes $\gtrsim$ the best-fit time scale for the radial basis function used in the regression. This is a consequence of under-sampling those larger bin sizes.

From these fits we obtain eclipse depth measurements of $680\pm68$ and $1081^{+54}_{-53}$ parts per million (ppm) in the 3.6 and 4.5 $\mu$m channels, respectively. These measurements account for the dilution factor of 1.207 in each band, and the 1--$\sigma$ ranges only overlap with the latter of the dilution-corrected depths of $1028\pm51$ and $1183\pm51$ ppm as reported in \citet{Deming2023}. Comparing with the existing sample of Hot Jupiters with \emph{Spitzer} eclipse depths, we borrow the scaling found in \citet{Garhart2019}, where eclipse depths are expressed as a ratio relative to the planet-to-stellar disk area ratio, in percent. At a planet-star radius ratio of $0.06128^{+0.00083}_{-0.00081}$, our eclipse depths can accordingly be re-expressed as $18.1\pm1.8\% \left(R_\mathrm{p}/R_\star\right)^2$ for the 3.6 $\mu$m channel and $28.8\pm1.4\% \left(R_\mathrm{p}/R_\star\right)^2$ for the 4.5 $\mu$m channel. The scaled depths at both wavelength ranges are consistent with the sequence of scaled depths across the sample of Hot Jupiters observed with warm \emph{Spitzer}, as shown in Figure 10 of \citet{Garhart2019}.

\subsection{Albedo, Heat Redistribution, and Temperatures}\label{sec:results:model}
\begin{deluxetable*}{lcccc}\label{table:eclipse_best-fit_parameters}
\tabletypesize{\footnotesize}
\tablewidth{0pt}
\tablecaption{Median and MLE parameter values for the free parameters in the thermal model fit to the eclipse depths of HD 202772 A b, for \emph{Spitzer} IRAC Channels 1 (centered at 3.6 $\mu$m) and 2 (centered at 4.5 $\mu$m).}
\tablehead{\colhead{Parameter} &
           \colhead{3.6 $\mu$m} &
           \colhead{4.5 $\mu$m}}
\startdata
Bond Albedo ($A_\mathrm{B}$)                & $0.42^{+0.20}_{-0.26}$ & $0.08^{+0.08}_{-0.06}$ \\
Heat Recirculation Factor ($\varepsilon$)   & $0.57^{+0.22}_{-0.34}$ & $0.09^{+0.09}_{-0.06}$ \\
Day-side Temperature ($T_\mathrm{d}$/K)     & $2130^{+102}_{-91}$    & $2611^{+46}_{-49}$     \\
Brightness Temperature ($T_\mathrm{b}$/K)   & $2313^{+109}_{-112}$   & $2750^{+80}_{-81}$     \\
\enddata
\end{deluxetable*}

\begin{deluxetable*}{lccccccc}\label{table:rad-con_parameters}
\tabletypesize{\scriptsize}
\tablewidth{0pt}
\tablecaption{Predicted observables from a set of 1-D climate models using the \texttt{PICASO} code \citep{Batalha2019,Mukherjee2023}, with model inputs from Table \ref{table:HD202772A_parameters} whose measurement uncertainties propagate to the uncertainties in our model observables. The model grid is described in \S \ref{sec:model:rad-con}. ``BB Deviation'' refers to the ``deviation from a blackbody'' metric as defined in \S 3.2 of \citet{Baxter2020}.}
\tablehead{ {} &
           \colhead{$\left[\mathrm{Fe/H}\right]$} &
           \colhead{3.6 $\mu$m Depth (ppm)} &
           \colhead{$T_{\mathrm{b}, 3.6}$ (K)} &
           \colhead{4.5 $\mu$m Depth (ppm)} &
           \colhead{$T_{\mathrm{b}, 4.5}$ (K)} &
           \colhead{BB Deviation (\%)} &
           \colhead{Bond Albedo ($A_\mathrm{B}$)}
           }
\startdata
\multirow{ 3 }{*}{$\frac{\mathrm{C}}{\mathrm{O}}/\odot = 0.5$} & $-1$ & $883^{+7}_{-6}$ & $2559^{+10}_{-9}$ & $1003^{+8}_{-10}$ & $2578^{+11}_{-14}$ & $0.0014^{+0.0014}_{-0.0017}$ & $0.26^{+0.04}_{-0.03}$ \\
                               &   0  & $828^{+6}_{-7}$ & $2479^{+9}_{-10}$ &  $981^{+8}_{-10}$ & $2547^{+11}_{-14}$ & $0.0047^{+0.0015}_{-0.0016}$ & $0.27\pm0.04$ \\
                               &   1  & $754^{+6}_{-5}$ & $2370^{+9}_{-7}$  &  $953^{+8}_{-9}$  & $2507^{+12}_{-13}$ & $0.0093^{+0.0013}_{-0.0015}$ & $0.26^{+0.04}_{-0.03}$ \\
\hline
\multirow{ 3 }{*}{$\frac{\mathrm{C}}{\mathrm{O}}/\odot = 1$}   & $-1$ & $882^{+6}_{-7}$ & $2557^{+9}_{-10}$ & $1010^{+8}_{-10}$ & $2588^{+11}_{-14}$ & $0.0022^{+0.0015}_{-0.0016}$ & $0.27^{+0.03}_{-0.04}$ \\
                               &   0  & $829\pm6$       & $2481\pm9$        & $1000^{+10}_{-8}$ & $2574^{+14}_{-11}$ & $0.0065^{+0.0016}_{-0.0014}$ & $0.26\pm0.04$ \\
                               &   1  & $752^{+5}_{-6}$ & $2367\pm7$        & $974\pm9$         & $2537^{+13}_{-16}$ & $0.0116^{+0.0015}_{-0.0014}$ & $0.26\pm0.04$ \\
\hline
\multirow{ 3 }{*}{$\frac{\mathrm{C}}{\mathrm{O}}/\odot = 2$}   & $-1$ & $\geq 855^{+6}_{-7}$\tablenotemark{*} & $2518^{+9}_{-10}$ & $\geq 1044^{+8}_{-10}$\tablenotemark{*} & $2637^{+11}_{-14}$ & $0.0083^{+0.0015}_{-0.0016}$ & $0.24^{+0.04}_{-0.03}$ \\
                               &   0  & $806^{+5}_{-7}$ & $2447^{+7}_{-10}$ & $1043^{+10}_{-9}$ & $2635^{+14}_{-13}$ & $0.0131^{+0.0017}_{-0.0014}$ & $0.26\pm0.03$ \\
                               &   1  & $714\pm5$       & $2311\pm8$        & $1030^{+10}_{-8}$ & $2617^{+14}_{-11}$ & $0.0211^{+0.0015}_{-0.0013}$ & $0.26\pm0.03$ \\
\enddata
\tablenotetext{*}{The thermal profile in this case exceeded the maximum temperature of the opacity data in the upper atmosphere. Therefore, these results should be considered lower limits.}
\end{deluxetable*}

\begin{figure}[htb!]
    \begin{center}
        \begin{tabular}{cc}
            \includegraphics[width=0.45\textwidth]{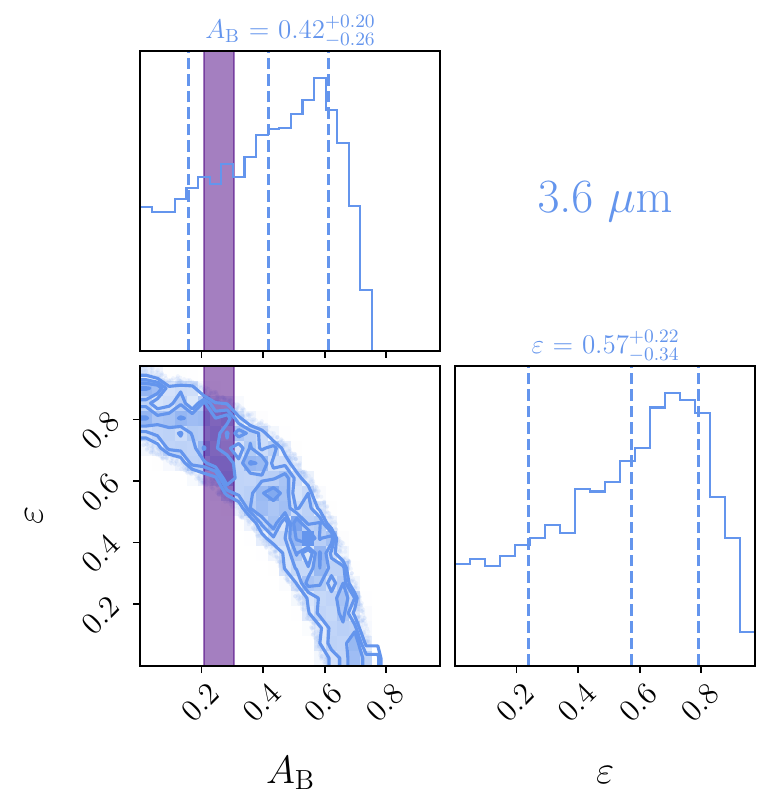} & \includegraphics[width=0.45\textwidth]{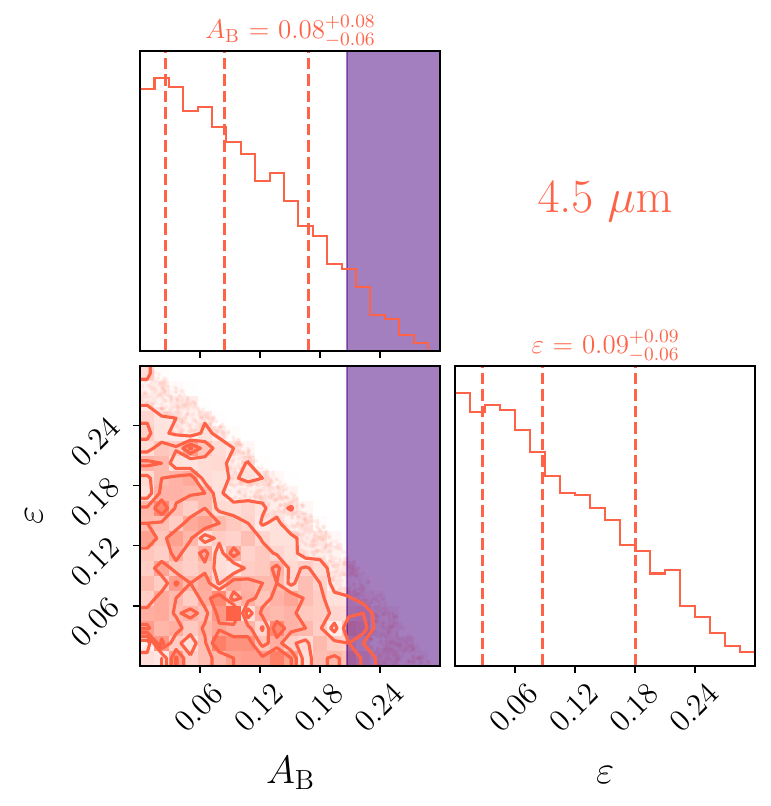} \\
        \end{tabular}
        \includegraphics[width=0.9\textwidth]{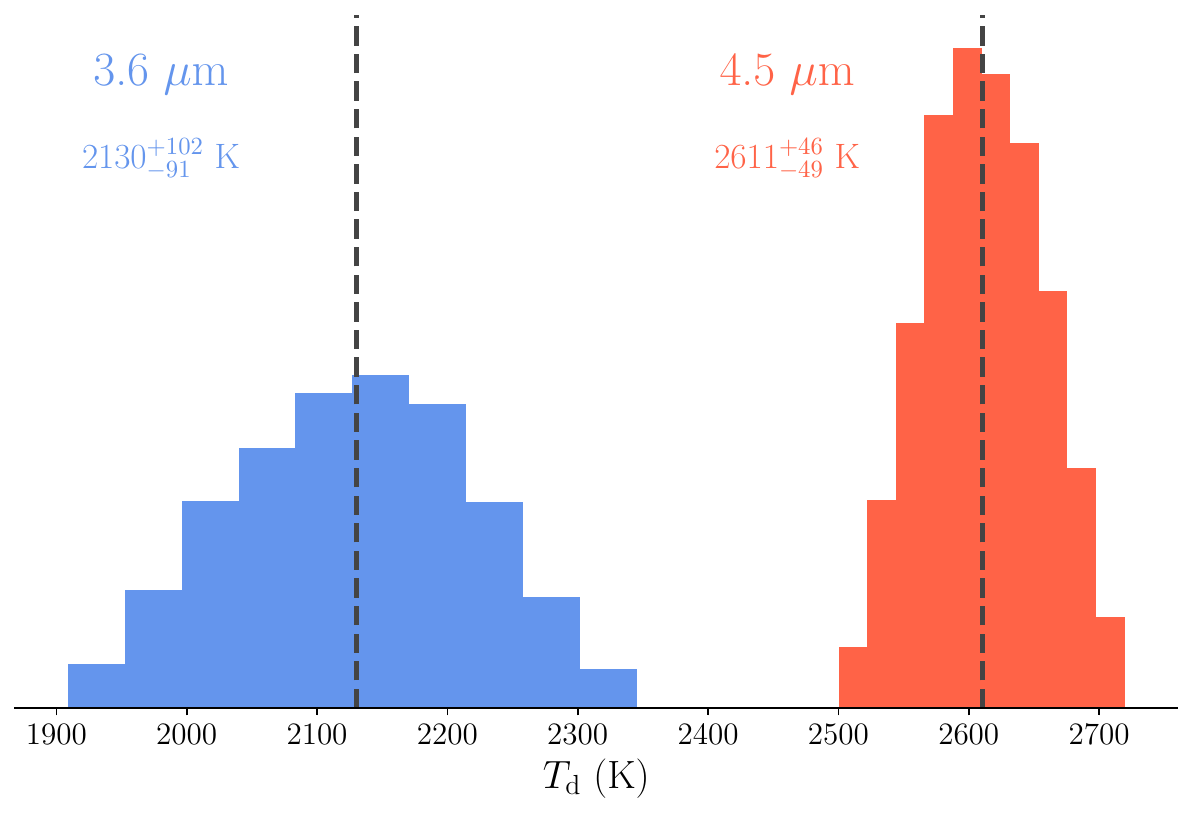}
    \end{center}
    \caption{Posterior distributions for the Bond albedo $A_\mathrm{B}$ and heat redistribution parameter $\varepsilon$ (top row of the corner plots, as defined in Equation \ref{eq:dayside_temperature} and Equation 4 of \citealt{Cowan2011}) which determine the day-side effective temperature for the 3.6 and 4.5 $\mu$m channels. The purple shaded regions show the range in Bond albedos predicted by the radiative-convective models described in \S \ref{sec:model:rad-con}. On the bottom row are the corresponding posterior distributions for the day-side temperature (as defined in Equation \ref{eq:dayside_temperature}).}
    \label{fig:albedo_eps_results}
\end{figure}

Our fits to the Bond albedo, heat redistribution efficiency, and day-side effective temperatures in each channel are shown in Figure \ref{fig:albedo_eps_results}, with results tabulated in Table \ref{table:eclipse_best-fit_parameters}. As expected, the distributions of the retrieved Bond albedos and heat redistribution efficiencies are highly correlated, with the distribution forming an arc which outlines the locus of points that generate similar day-side temperatures. This is expected; the use of the parameter estimation here is to estimate the propagated range of uncertainties in each parameter, rather than to provide any disentangling of the inherently degenerate parameters. The distributions overlap considerably, with the lower 95\% confidence interval of each being consistent with zero. This is also true of the heat redistribution efficiency, suggesting a day side that is absorbing most of the incident stellar radiation and keeping it from circulating to the night side.

To provide a theoretical constraint on the Bond albedo, we use the results from our set of 1-D radiative-convective climate models. Table \ref{table:rad-con_parameters} shows the expected eclipse depths in each \emph{Spitzer} band, as well as the inferred brightness temperatures. Additionally, one can obtain an estimate for the albedo via the effective temperature: since the model is set up to calculate a 1-D emission spectrum at the limiting case of a tidally-locked day side, this is equivalent to neglecting heat redistribution. Therefore, we can directly estimate $A_\mathrm{B}$ by setting $\varepsilon \rightarrow 0$ in Equation \ref{eq:dayside_temperature}. The generated models span a range of 0.5 to 2 in C/O ratio relative to solar and $-1$ to 1 in log metallicity; the Bond albedo only varies from $\approx 0.21$--0.31, with the range over-plotted on the corner plots in Figure \ref{fig:albedo_eps_results}. However, we find that the model case returning the deepest eclipse depths --- that of the case C/O = $2\mathrm{C/O}_\odot$, $\left[\mathrm{Fe/H}\right]=-1$ --- returns a T--P profile that hits the upper limit of temperatures for the available opacity data. This means that we are not able to accurately estimate the entire thermal contribution from the uppermost layers of the model atmosphere, where a thermal inversion may be expected for such a strongly irradiated planet. Therefore, we consider the results an upper limit for this case. The effective temperature histograms show the limited range in model day-side effective temperatures as well, at 2521--2543 K. With the model Bond albedos, one can estimate the degree to which thermal emission must be affected by heat redistribution or other effects to obtain the observed eclipse depths in each band. Without eclipse observations in the optical, redistribution efficiency may be degenerate with a variety of atmospheric effects, including opacity from clouds; nevertheless, the heat redistribution parameter stands in as a first-order accounting of whichever effect is at play. Within the band of model albedos, we find the 68\% range of $\varepsilon$ is $0.71\pm0.10$ at 3.6 $\mu$m, and $0.03^{+0.03}_{-0.02}$ at 4.5 $\mu$m (Figure \ref{fig:eps_from_radcon}). That is, the implied heat flow from the day to night sides is much stronger at 3.6 $\mu$mm than at 4.5 $\mu$m, the latter of which implies almost no night-side flow.

\begin{figure}[htb!]
    \begin{center}
        \includegraphics[width=\textwidth]{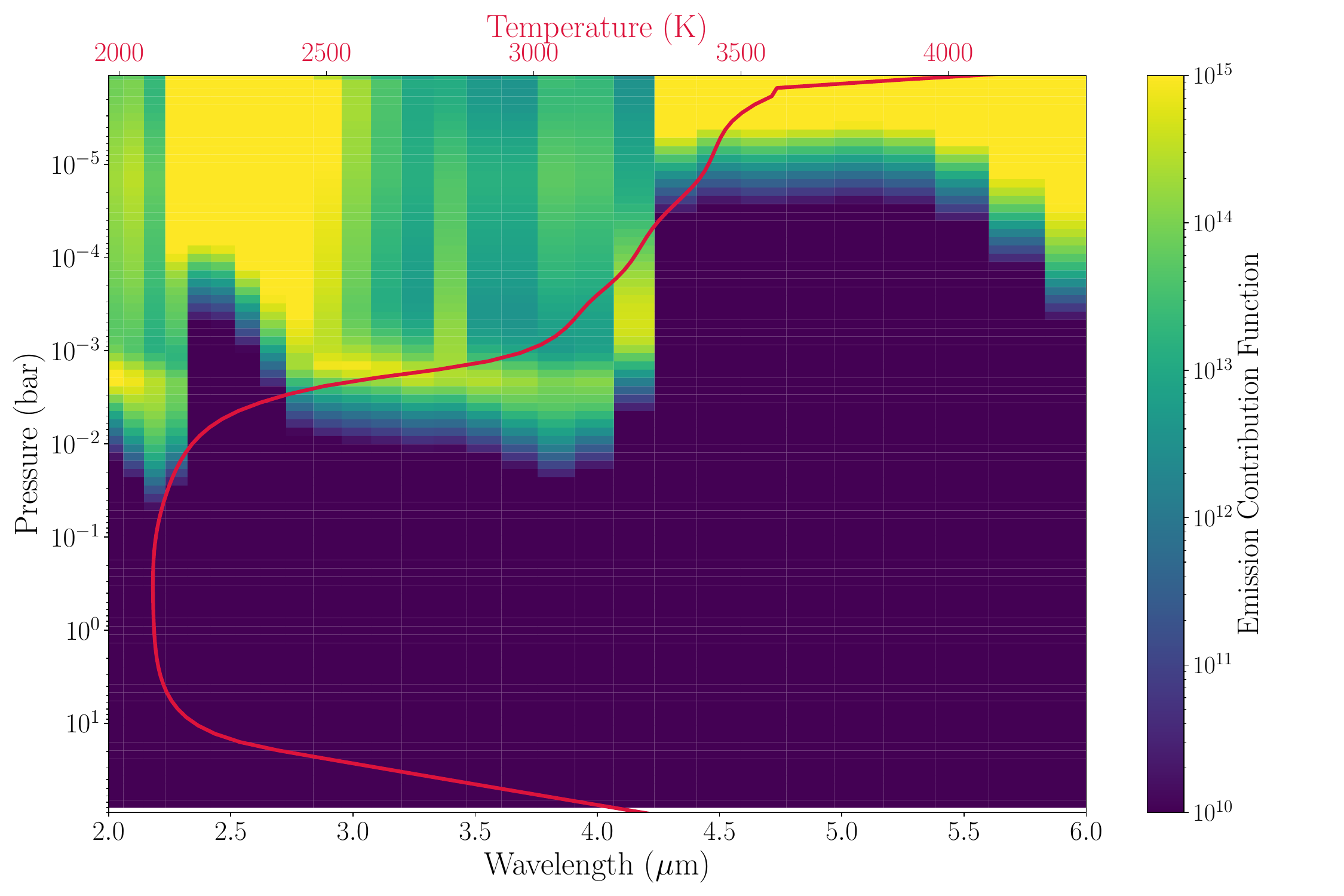} \\
    \end{center}
    \caption{The T-P profile and emission contribution function calculated by the radiative-convective model described in \S \ref{sec:model:rad-con}, for the case of C/O = $2\times$solar, $\left[\mathrm{Fe/H}\right]=1$, which is nominally closest to the measured eclipse depths. However, our constraints on the albedo and heat redistribution efficiencies imply that the 3.6 $\mu$m eclipse data are not well described by a 1-D dayside-only model that does not account for heat flow to the night side.}
    \label{fig:radcon_results}
\end{figure}

\begin{figure*}[htb!]
\begin{center}
\includegraphics[width=8.5cm]{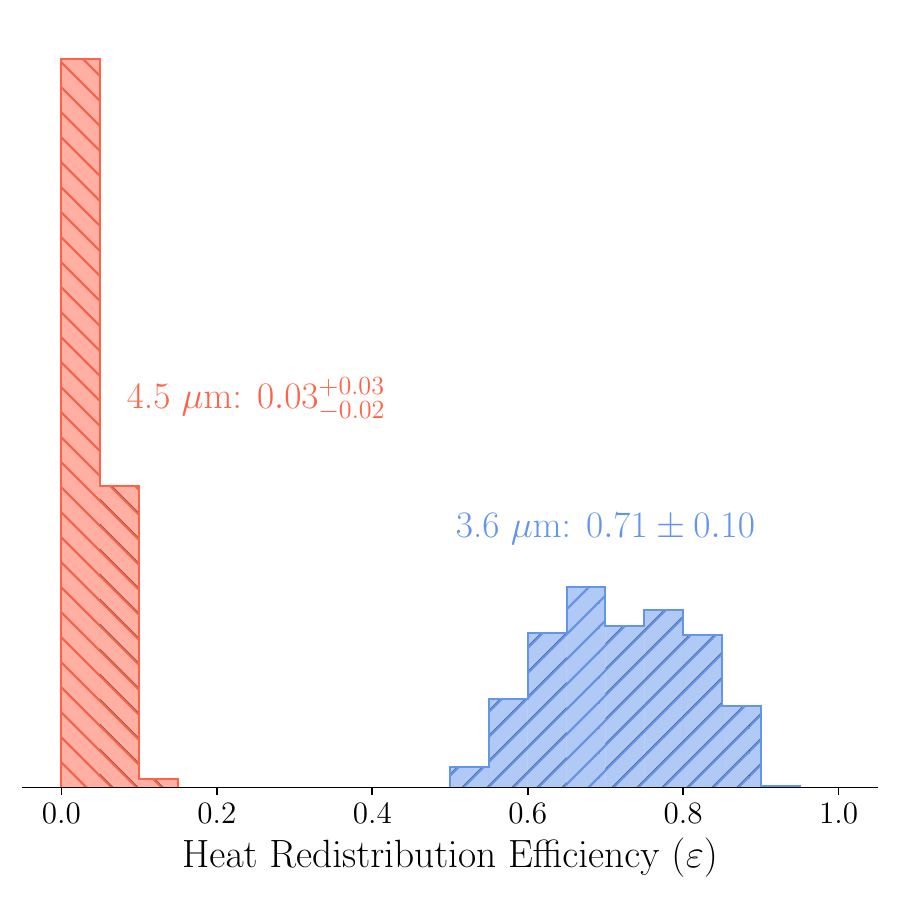}
\caption{Distributions of heat redistribution efficiency values ($\varepsilon$), as calculated from the nested sampling results, that fall within the range of Bond albedos derived from the effective temperatures of the 1-D radiative-convective climate models. The values from the 3.6 $\mu$m channel are in light blue, and those from the 4.5 $\mu$m channel are in red. The two distributions are distinct, with a distribution for the 3.6 $\mu$m channel that suggests a heat redistribution much more even (i.e.~much closer to 1) than that for the 4.5 $\mu$m channel which suggests virtually no heat flow to the night side.}
\label{fig:eps_from_radcon}
\end{center}
\end{figure*}

\section{Discussion} \label{sec:discussion}
\subsection{Day-side Temperatures versus Bond Albedos}\label{sec:discussion:Td_vs_AB}
\begin{figure*}[htb!]
\begin{center}
\includegraphics[width=8.5cm]{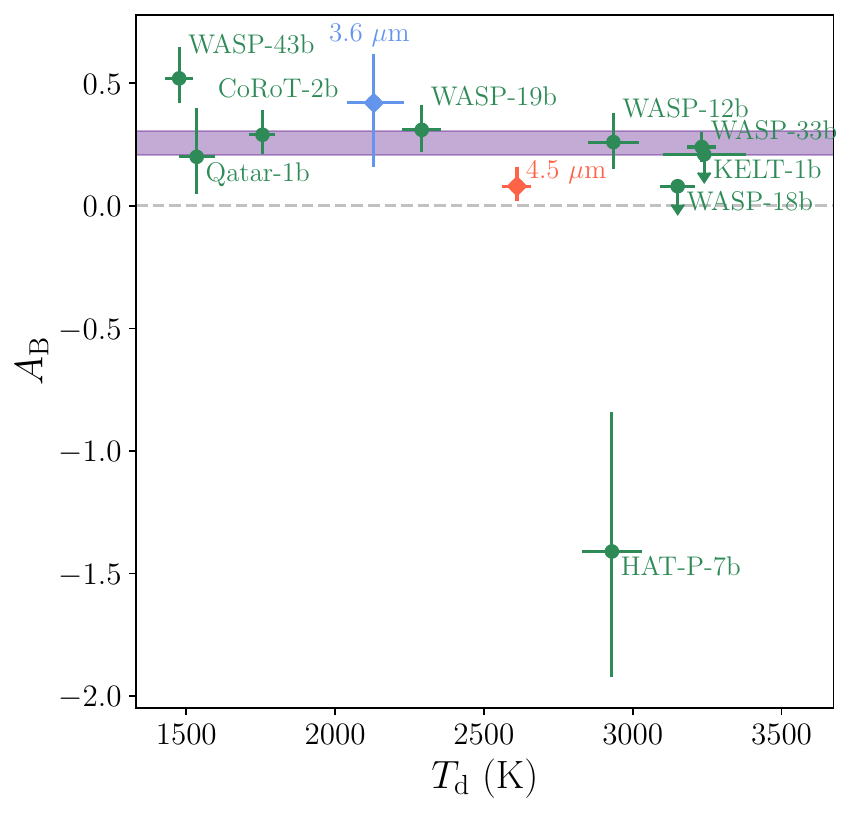}
\caption{The day-side temperatures $T_\mathrm{d}$ versus Bond albedos $A_\mathrm{B}$ for HD 202772 A b in the 3.6 $\mu$m (light blue) and 4.5 $\mu$m (red). The purple shaded region is the range of Bond albedos calculated from the radiative-convective models described in \S \ref{sec:model:rad-con}, and listed in Table \ref{table:rad-con_parameters}. Also plotted in green are the same quantities for other \emph{TESS}-observed Hot Jupiters, calculated from their \emph{Spitzer} 4.5 $\mu$m phase curves as analyzed in \citet{Bell2021}, listed in Table 8 of \citet{Wong2021}. The calculated Bond albedos for HD 202772 A b are consistent with the generally flat behavior of close-in giants with day-side temperature, with the notable exception of HAT-P-7 b, whose negative albedo is due to it having a significantly brighter day side than would be expected from purely thermal emission.}
\label{fig:Td_vs_AB}
\end{center}
\end{figure*}

Works such as \citet{Wong2021} have noted a positive correlation in geometric albedo with day-side temperature. Geometric albedos depend both on the wavelength and the orbital phase angle, and are usually reported at zero phase angle (secondary eclipse). They are typically measured in the optical where the contribution of reflected light is relatively high. This trend reflects a trend in the cloud composition and distribution with temperature. The Bond albedo, in contrast, is calculated by integrating over all orbital phases and wavelengths, and is therefore more sensitive to the thermal emission of the planet than the geometric albedo. We cannot directly calculate geometric albedo without a publication on secondary eclipses in the optical, therefore we are limited to comparing our results with the thermally-derived Bond albedos. The Bond albedos show very little (or a very weak negative) correlation with day-side temperature, as illustrated in Figure \ref{fig:Td_vs_AB}. The phase integral, typically defined as the ratio of the Bond to geometric albedos ($q \equiv A_\mathrm{B}/A_\mathrm{g}$), can be calculated for planets with both measurements, and shows a negative correlation with $A_\mathrm{g}$ (see Figure 12 in \citealp{Wong2021}). This renders the relation between $A_\mathrm{B}$ and $T_\mathrm{d}$ very nearly flat, with HAT-P-7b as a notable exception (see the caption in Figure \ref{fig:Td_vs_AB}).

\subsection{Brightness Temperature Ratios}\label{sec:discussion:Tb}
Our ratio of $T_{\mathrm{b}, 4.5}/T_{\mathrm{b}, 3.6} = 1.19^{+0.10}_{-0.09}$ at $T_\mathrm{eq} = 2132$ K, as calculated from the medians of our posterior distributions, is consistent with the existing sample of planets observed in eclipse with warm \emph{Spitzer}, as presented in Figures 19 and 21 of \citet{Garhart2019}, as well as Figures 5 and 6 of \citet{Wallack2021} and Figure 7 of \citet{Deming2023}. Because our 3.6 $\mu$m eclipse depth is lower than that obtained in \citet{Deming2023}, we infer a brightness temperature ratio on the high end of the existing sample. Calculating the ``deviation from blackbody'' metric as defined in \citet{Baxter2020}, we obtain an excess of $0.026\pm0.005$\%, which fits well within the range of values spanned by the hot Jupiter sample in Table 1 of \citet{Baxter2020}, and is comparable to the excesses obtained with hot Jupiters at similar equilibrium temperatures. This suggests that the fact that HD 202772 has evolved off the main sequence has not imparted any significant changes to this planet's atmosphere, at least in eclipse photometry. Comparing with the predicted blackbody deviations for the radiative-convective models, we see that the observed excess is most consistent with the model with twice the solar C/O ratio and a solar metallicity. This model is also most consistent with the measured brightness temperatures themselves. However, within uncertainties the measurements are also at least marginally consistent with all except the low-metallicity cases ($0.1\times$solar) for sub-solar and solar C/O. In other words, either a super-solar metallicity or super-solar C/O ratio is preferred for our models to reproduce the observed differences in brightness temperatures between the \emph{Spitzer} bands. This result is also consistent with the finding of \citet{Deming2023} that the sample of hot Jupiter eclipse depths falls within the tracks of their atmospheric models bounded by solar and $30\times$ solar metallicities, though they report that higher metallicities are more often consistent with planets with temperatures $\lesssim 1200$ K. In the absence of additional observations at higher spectral resolution and signal-to-noise, however, we lack the ability to comment with much more detail on the state and/or evolutionary history of HD 202772 A b's atmosphere.

\section{Conclusions} \label{sec:conclusions}
We present a reduction of 3.6 and 4.5 $\mu$m \emph{Spitzer} photometry of the HD 202772 A (a.k.a.~TOI-123) system during individual secondary eclipses of the planetary-mass companion HD 202772 A b. Isolating the eclipse signal via pixel-level decorrelation methods yields 68\% confidence intervals of $680\pm68$ and $1081^{+54}_{-53}$ ppm at 3.6 and 4.5 $\mu$m respectively. These correspond to day-side temperatures of $2130^{+102}_{-91}$ and $2611^{+46}_{-49}$ K, respectively, compared with the estimated equilibrium temperature of $2132^{+28}_{-30}$ K from \citet{Wang2019a}. The 68\% confidence intervals of Bond albedos $A_\mathrm{B}$ and heat redistribution efficiencies $\varepsilon$ are $0.42^{+0.20}_{-0.26}$ and $0.57^{+0.22}_{-0.34}$ respectively for 3.6 $\mu$m, and $0.08^{+0.08}_{-0.06}$ and $0.09^{+0.09}_{-0.06}$ for 4.5 $\mu$m. When considering the range of Bond albedos consistent with our set of 1-D radiative-convective atmospheric models, the ranges of $\varepsilon$ are distinct, at $0.71\pm0.10$ and $0.03^{+0.03}_{-0.02}$ for 3.6 and 4.5 $\mu$m respectively. Based on these, we find that
\begin{itemize}
    \item The Bond albedos are consistent with the trend in albedos with day-side effective temperatures, comparing with the samples presented in \citet{Bell2021} and \citet{Wong2021}.
    \item The inferred brightness temperatures are $2313^{+109}_{-112}$ and $2750^{+80}_{-81}$ K for 3.6 and 4.5 $\mu$m, respectively, for this planet are high compared with those in the samples presented in works such as \citet{Garhart2019,Baxter2020,Wallack2021}; and \citet{Deming2023}.
    \item The blackbody deviation metric of HD 202772 A b, as originally defined in \citet{Baxter2020}, is consistent with the existing characterized population of hot Jupiter hosts, and is most consistent with our radiative-convective model that uses solar metallicity and a C/O twice that of the solar value.
\end{itemize}

In the discovery paper for HD 202772 A b, \citet{Wang2019a} briefly discuss the opportunity for follow-up observation of the planet in transit using the transmission spectrum metric, which we estimated earlier as $\approx 210$ if the metric as defined in \citet{Kempton2018} is used, and the associated scale factor as calculated from \citet{Louie2018} is extended beyond 1 Jupiter radius. The metric is designed as the estimate of the S/N one may expect if the planet were observed in transit in a 10-hour observing program with the Near-Infrared Slitless Spectrograph (NIRISS) instrument on \emph{JWST}. Though we are applying the metric to a planet with a super-Jupiter radius and a target star brighter than what was used in its original construction, the results still imply that HD 202772 A b competes with the hottest and brightest Hot Jupiters observed with \emph{TESS} in terms of expected S/N in observation with \emph{JWST}. At this temperature range, we will expect to place much more precise constraints on the thermal structure of the atmosphere with an instrument such as NIRISS, as well as detections of cloud opacity from species such as silicates, quartz, and iron which would be expected to contribute to the emission and transmission spectra \citep[as predicted in works such as][]{Marley2013,Helling2019c,Gao2020}. This planet is also a great candidate for spectra into the mid-infrared ($\sim 10$ $\mu$m), as the combination of near- and mid-infrared at \emph{JWST} resolutions can constrain cloud opacity from molecules such as quartz, iron, and silicates which are expected to be the primary possible condensates for planets in this temperature range. As an inflated Hot Jupiter orbiting an evolved early-type star in a binary system, characterizing its atmospheric composition and structure is key to untangling the formation and evolution of this system.

\begin{acknowledgements}
T.F. acknowledges support from the University of California President's Postdoctoral Fellowship Program. K.B. acknowledges support from NASA Habitable Worlds grant No. 80NSSC20K1529. P.D. acknowledges support by a 51 Pegasi b Postdoctoral Fellowship from the Heising-Simons Foundation and by a National Science Foundation (NSF) Astronomy and Astrophysics Postdoctoral Fellowship under award AST-1903811.
\end{acknowledgements}

\bibliography{library}{}
\bibliographystyle{aasjournal}

\software{astropy \citep{ast13,Bradley2023}, dynesty \citep{Speagle2020}, emcee \citep{Foreman-Mackey2013}, matplotlib \citep{hun07}, numpy \citep{van11}, PICASO \citep{Batalha2019,Marley2021,Mukherjee2023}, synphot/pysynphot \citep{pysynphot,synphot}}


\end{document}